\documentclass[9pt,twocolumn,twoside]{opticajnl}
\journal{opticajournal} 

\setboolean{shortarticle}{true}

\usepackage{lineno}

\title{Driving Spintronic Terahertz Emitters at GHz Repetition Rates}

\author[1,2]{Thomas A. Gething}
\author[1,2]{Henry De Libero}
\author[1,3]{Morgan T. Hibberd}
\author[2]{Thomas Thomson}
\author[2]{Paul W. Nutter}
\author[1,3]{Darren M. Graham}

\affil[1]{Department of Physics and Astronomy \& Photon Science Institute, The University of Manchester, Manchester M13 9PL, United Kingdom}
\affil[2]{Nano Engineering and Spintronic Technologies Group,
Department of Computer Science, The University of Manchester, Manchester M13 9PL, United Kingdom}
\affil[3]{The Cockcroft Institute, Sci-Tech Daresbury, Keckwick Lane, Warrington WA4 4AD, United Kingdom}

\affil[*]{thomas.gething@manchester.ac.uk}

\begin{abstract}
Spintronic terahertz emitters (STEs) are versatile sources of terahertz (THz) frequency radiation that offer high electric field strengths and gap-free bandwidths covering the THz (0.1-30\,THz) band. In addition, they are low-cost to produce and easy to use in a variety of optical setups. However, an obstacle to their wider applicability is the recently reported decrease in optical fluence damage threshold when driven with high-repetition-rate lasers. Here, we use an asynchronous optical sampling (ASOPS) THz time-domain spectrometer (THz-TDS) to observe the degradation of the THz signal in real-time when driven by a 1\,GHz repetition-rate laser. We demonstrate that the loss of THz emission is permanent and is likely the result of thermal damage to the STE structure. However, by utilizing substrates with high thermal conductivities, an increase of the damage threshold of over three orders of magnitude is observed enabling stable emission from a STE for the first time when driven by a high-repetition-rate laser. 
\end{abstract}

\setboolean{displaycopyright}{false} 

\begin{document}

\maketitle
Since spintronic terahertz emitters (STEs) were first introduced by Kampfrath \textit{et al.} \cite{Kampfrath2013} they have been shown to produce terahertz (THz) pulses with high electric-field strengths (up to 1.95\,MV\,cm$^{-1}$ \cite{Huang2025}), a gap-free spectral emission, and bandwidths of up to 30\,THz.\cite{Rouzegar2023,Bull2021,Seifert2016,Seifert2017} They consist of nanometer thin layers of ferromagnetic (FM) and heavy metals (HM) and can be easily fabricated with established thin-film deposition methods \cite{Bull2021}. The operation of STEs relies on spin-to-charge conversion, via the inverse spin-Hall effect, to convert a femtosecond laser induced spin-polarized current into a transient charge current that acts as a dipole source of THz radiation. In addition, STEs can be easily integrated into THz time-domain spectrometers (THz-TDS) as they can be driven by pulse energies as low as nanojoules \cite{Nandi2019}, with the polarization of the emitted THz being entirely dependent on the in-plane magnetization direction of the FM layer. \cite{Hibberd2019} STEs are typically exploited on THz-TDS systems with driving laser repetition rates of up to 80\,MHz, however, a move to higher repetition-rate lasers could lead to  improved signal-to-noise ratios and thus shorter acquisition times \cite{Millon2023, Beck2019}, potentially unlocking broader adoption of these versatile THz radiation emitters. 

In 2023, Paries \textit{et al.}\cite{Paries2023} showed that STEs grown on the end of an optical fiber can be driven using a laser with a repetition rate of 100\,MHz, reporting a damage threshold of the STE at pulse energies less than 20\,$\mu$J\,cm$^{-2}$. However, this is in stark contrast to the 4-8\,mJ\,cm$^{-2}$ typically reported for STEs driven at lower repetition-rate lasers (between 1\,kHz and 400\,kHz).\cite{kumar2021,Vogel2022} Paries \textit{et al.}\cite{Paries2024} further explored optical damage thresholds of fiber-tip STEs at high laser repetition rates (between 200\,kHz and 1\,GHz). Above a repetition rate of 4\,MHz, they observed a fluence threshold that decreased linearly with increasing repetition rate, which corresponds to a continuous-wave laser power density of 4.8\,kW\,cm$^{-2}$. This damage was attributed to irreversible changes of the film composition resulting from heating of the STEs by the laser irradiation. At the highest repetition rates investigated (1\,GHz), the fluence threshold was 4.8\,$\mu$J\,cm$^{-2}$, where no observable THz emission was reported. 

Here, we demonstrate for the first time THz emission from a STE driven by a high (GHz) repetition rate laser. Consistent THz emission was achieved by careful selection of substrate materials with sufficient thermal conductivities to control the thermal distribution in the STE and to limit the onset of thermal damage. This was made possible by the use of real-time, direct monitoring of the degradation of the THz emission signal using asynchronous optical sampling (ASOPS) THz-TDS.

\begin{figure}[h!]
\centering\includegraphics[width=\linewidth]{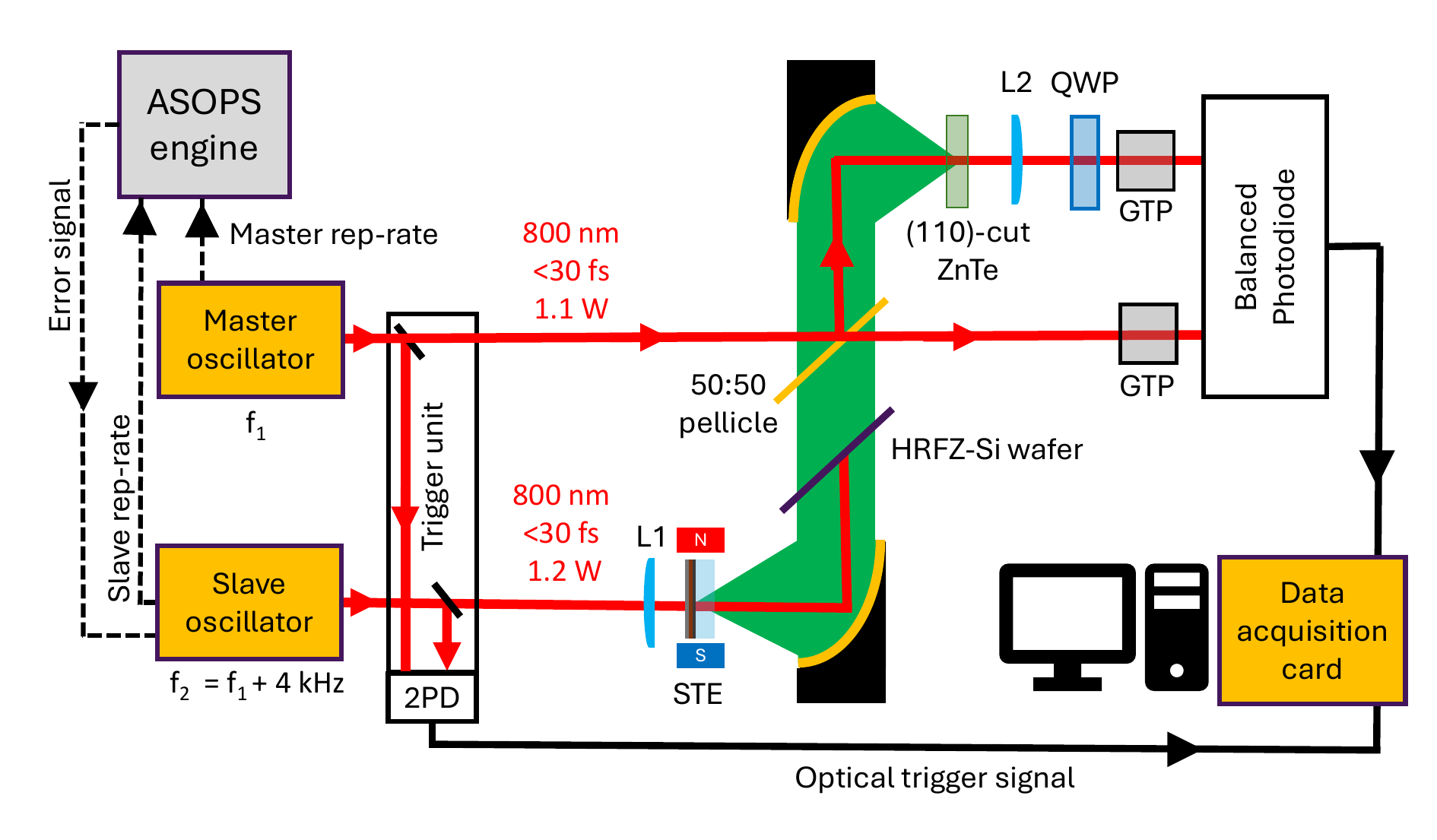}
\caption{Schematic diagram of the ASOPS-based THz-TDS system incorporating two laser oscillators (Taccor power x6, Novanta) at approximately 1\,GHz repetition rates and 1.2\, W average power, frequency offset by synchronisation electronics (TL1000 ASOPS engine, Novanta). The HRFZ-Si wafer beam block was placed at Brewsters angle to minimize THz reflection losses. GTP: Glan-Taylor polariser. QWP: Quarter-wave plate. L1/L2: Lenses. 2PD: two-photon detector. }
\label{fig:figure_1}
\end{figure}

\textbf{Experimental methods.} A schematic diagram of the ASOPS-based THz-TDS is shown in Fig. \ref{fig:figure_1}. The ASOPS technique replaces the slow mechanical delay stage typically used in THz-TDS measurements with two femtosecond lasers, of different  repetition rates ($f_1$ and $f_2$), offset by $\Delta f$. The laser repetition rates were stabilized by active repetition-rate offset locking in a master-slave configuration. \cite{Gebs2010,Klatt2011} The slave laser was focused using a 200\,mm focal length lens onto a STE to generate the emitted THz radiation, which was collected by a 50.8\,mm-diameter gold-coated $90^{\circ}$ off-axis parabolic mirror (OAPM) with a focal length of 50.8\,mm. A high-resistivity float-zone silicon (HRFZ-Si) wafer was placed close to Brewster's angle for the horizontally polarized THz radiation to block any residual light from the driving laser from propagating to the detection optics. A 45:55 pellicle beam splitter enabled co-propagation of the laser probe beam and the THz beam to the detection optics. The THz beam and the probe beam were then focused onto a (110)-cut ZnTe detection crystal with a thickness of 2\,mm using a second $90^{\circ}$ 50.8\,mm diameter OAPM with a focal length of 50.8\,mm. Electro-optic sampling was used to measure the THz electric-field using the crossed-polarizer configuration \cite{Jiang1999}. Using a repetition-rate offset of 4\,kHz enabled a single THz waveform to be acquired in a real time of 250\,$\mu$s (1/$\Delta f$) over a time-delay range of 1\,ns (1/$f_1$).

To determine the excitation laser fluence, the spot size at the position of the STE was measured using the knife-edge method to obtain intensity profiles and the average power was measured. The definition of fluence used here was the average power divided by the $1/e^2$ beam area. The measured intensity profiles for all excitation spots used are included in Supplemental 1 (Section 1, Fig. S1.). As the nature of the damage was reported as thermal \cite{Paries2024}, a thermal camera (FLIX-C3X) was used to obtain images of the change in temperature of the STEs.

The STEs used in this study were fabricated by magnetron sputtering and consisted of trilayer films of W/CoFeB/Pt with nominal layer thicknesses of 2\,nm deposited on a variety of substrates. After deposition, x-ray reflectometry (XRR) measurements were performed to confirm the layer thicknesses of the STEs grown. Further details of the deposition and the XRR results are available in Supplemental 1 (Section 2). To saturate the magnetisation of the CoFeB layer of the STE, two permanent neodymium (NdFeB) magnets were used that produced a magnetic field of 47\,mT at the center of the STE. \cite{Rouzegar2023}

Two sets of STEs were produced for this study. The first set was used to investigate the temporal degradation and thermal properties of STEs and were grown on a $20\times20\times0.5$\,mm fused silica substrate and a $10\times10\times1$\,mm sapphire substrate (single-side polished). THz waveforms were obtained sequentially over the course of half an hour with one minute acquisition times. After each THz waveform acquisition, the drive beam was blocked and the STE was left to return to thermal equilibrium, which allowed the STEs to recover any non-permanent changes in THz emission due to any increase in temperature. After this, a subsequent THz waveform was obtained to determine if the THz waveform amplitude had returned to the initial value and thus identify whether any damage had occurred.

The second set of STEs were used to further investigate the impact of substrate thermal conductivity on the damage threshold of STEs and were deposited on five different substrates: fused silica, sapphire, MgO (all $10\times10\times0.5$\,mm), diamond ($10\times10\times0.4$\,mm) and commercially available high-resistivity float-zone silicon with a high reflectivity dielectric mirror coating (circular, of diameter $12$\,mm and thickness of $0.66$\,mm). 

To assess the damage threshold of the second set of STEs, the relationship between excitation fluence and peak-to-peak emission amplitude was measured. The incident average drive laser power, and thus fluence, was controlled using a half-wave plate mounted in a motorized rotation mount and a fixed Glan-Taylor linear polarizer allowing for repeatable ramping of the excitation fluence. As damage was evident in less than ten minutes, THz waveforms with this acquisition time were measured. This acted as both a measure of STE performance and as an exposure stage to cause damage. After each exposure, the driving power was reduced to the previous power and another THz waveform was acquired. Any decrease in the measured amplitude before and after exposure to the higher driving laser power indicates whether damage had occurred. For an example of this approach see Supplemental 1 (Section 3, Fig. S3.).

\begin{figure}[h!]
\centering\includegraphics[width=\linewidth]{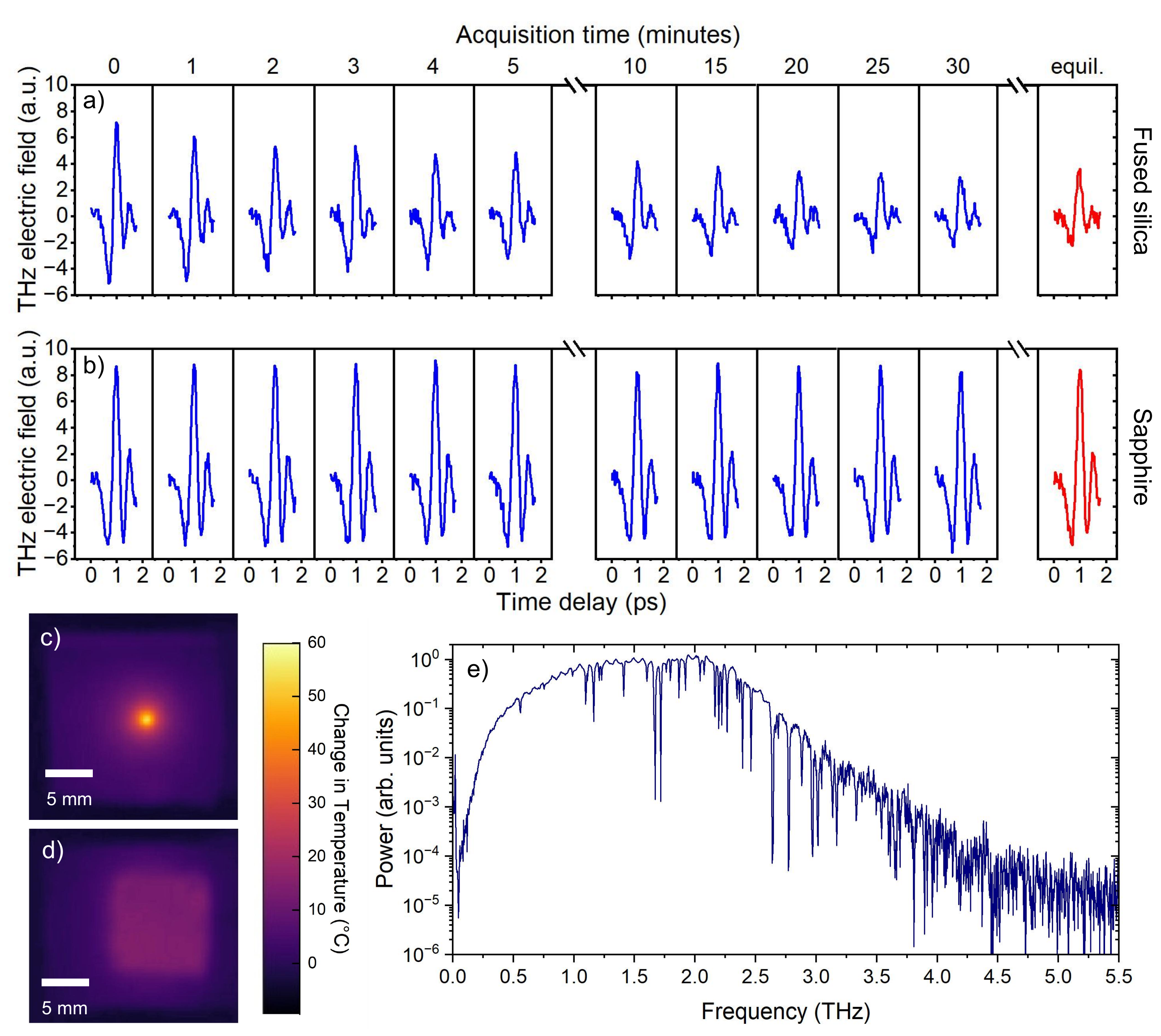}
\caption{Series of measured THz waveforms from STEs grown on fused silica (a) and sapphire (b) when driven with a fluence of 0.97\,$\mu$m\,cm$^{-2}$. The equilibrium (equil.) waveform was taken after blocking the drive laser and allowing the STE to return to room temperature. c) and d) show the thermal distribution of the two STEs after 30 minutes exposure to the driving laser. e) Power spectrum of the STE grown on sapphire with a 48 minute acquisition time enabled by the THz emission stability due to the use of the sapphire substrate.}
\label{fig:FSSP_time_and_temp}
\end{figure}

\textbf{Results and discussion.} The excitation laser spot size for the first set of STEs was found to be elliptical with major and minor axes of $0.300\pm0.007$\,mm and $0.170\pm0.007$\,mm respectively. The average drive laser power was 388\,mW, giving an excitation fluence of $0.97\,\mu$J\,cm$^{-2}$. Figure \ref{fig:FSSP_time_and_temp}a) shows the THz waveforms measured from the STE grown on fused silica. The waveforms reveal a decay in the THz amplitude across subsequent acquisitions, decaying by 45\% of the initial peak-to-peak amplitude after approximately 15 minutes of laser exposure time. This drop in emission amplitude is unexpected as the fluence used here is five times less than the reported damage threshold for glass fiber-tip STEs. \cite{Paries2024} In contrast, pulses acquired using the STE grown on sapphire, Fig. \ref{fig:FSSP_time_and_temp}b), show no obvious deterioration in pulse amplitude over the same time period (a decrease of 1\% is within the measurement noise), implying that the choice of substrate may prevent degradation of emitter performance. This contrasts to the reported behavior of fiber-tip STEs where the damage threshold for fused silica and sapphire are shown to be similar. \cite{Paries2023}

 Figure \ref{fig:FSSP_time_and_temp}c) illustrates the temperature distribution across the STE on fused silica, where it is clear that the heat is confined to a region close to the drive laser incidence point. However, in the case of the STE on sapphire, Fig. \ref{fig:FSSP_time_and_temp}d) the heat is seen to be more evenly distributed across the emitter surface.
 
 The consistent amplitude of the emitted THz pulses from the STE on sapphire, along with the thermal image in Figure \ref{fig:FSSP_time_and_temp}c), suggests an improvement in heat dissipation, likely to be due to higher thermal conductivity in sapphire (between 33\,W\,m$^{-1}$\,K$^{-1}$ and 43\,W\,m$^{-1}$\,K$^{-1}$ at room temperature) \cite{Sp_K,Qz_Sp_K,MgO_Sp_k} compared to that of fused silica (approximately 1.15\,W\,m$^{-1}$\,K$^{-1}$ \cite{FS_K,Qz_FS_K}. This observation suggests that using a substrate with a higher thermal conductivity may prevent deterioration of the THz emission from the STE due to improved heat dissipation across the STE.  The discrepancy with the report by Paries \textit{et al.} of sapphire and glass fiber-tip STEs having similar damage thresholds may be due to the reduced volume in a fiber, which allows the heat to spread laterally in this case. Figure \ref{fig:FSSP_time_and_temp}e) shows the power spectrum obtained from the STE grown on sapphire with a long acquisition time enabled by the improved emission stability. 

\begin{figure}[h!]
\centering\includegraphics[width=\linewidth]{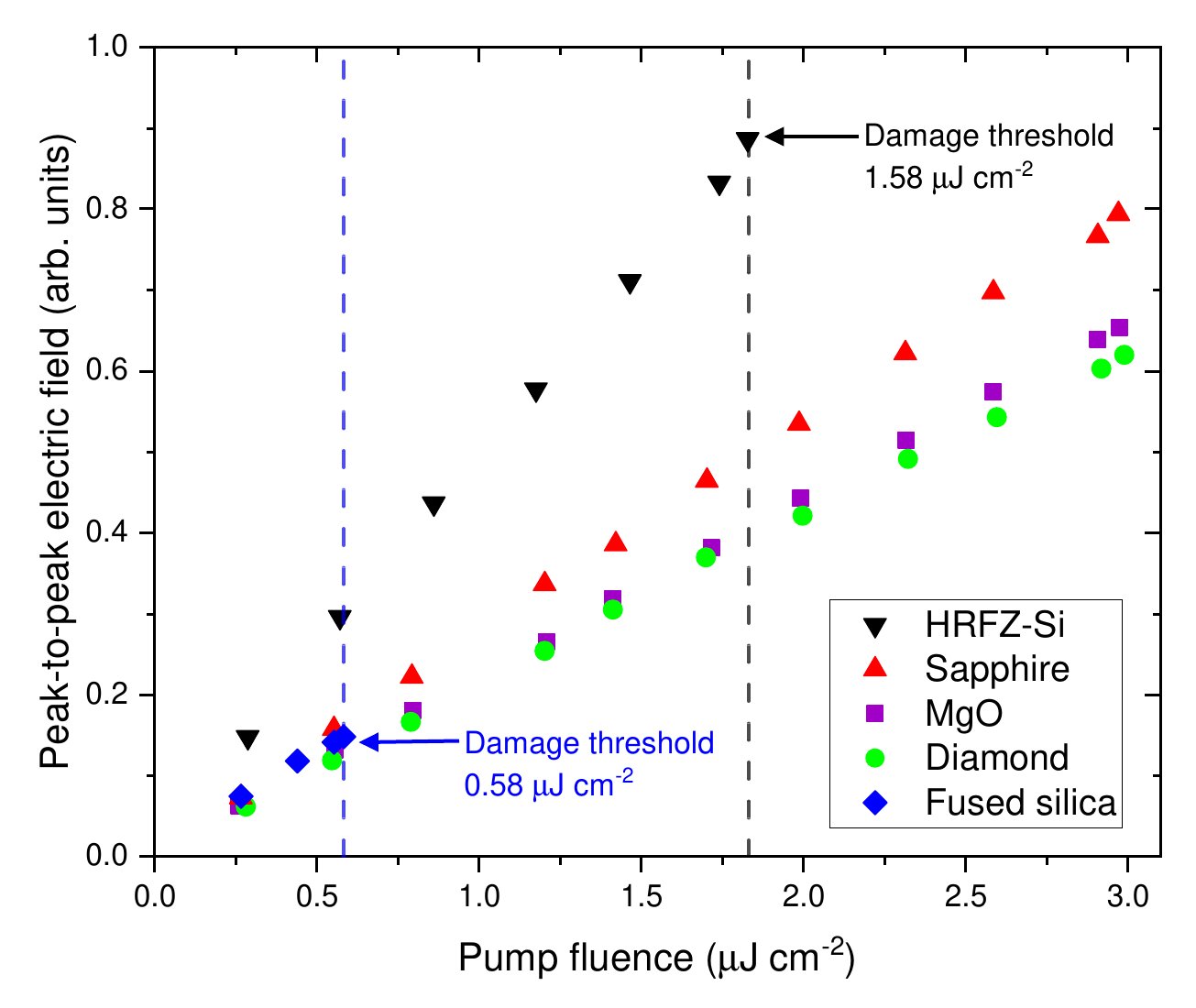}
\caption{Measured peak-to-peak THz electric field as a function of drive beam fluence for STEs grown on five substrate materials: HRFZ-Si, sapphire, MgO, diamond and fused silica. Measurements were taken over a low fluence range which was achieved using a spot size of $(248.6\pm0.5\,\times\,175.0\pm0.5)\,\mu$m.}
\label{fig:large_p2p}
\end{figure}

Figure \ref{fig:large_p2p} shows the peak-to-peak THz electric field extracted from the THz waveforms plotted against excitation fluence for the second set of STEs when driven by a beam with dimensions of $(248.6\pm0.5\,\times\,175.0\pm0.5)\,\mu$m. The highest fluence used to drive the STE on fused silica, where damage of the STE was not observed, was $0.58\pm0.03\,\mu$J~cm$^{-2}$ and for HRFZ-Si it was $1.83\pm0.01\,\mu$J~cm$^{-2}$. No damage was observed for STEs grown on MgO, diamond and sapphire for the fluence range investigated at this driving beam spot size, implying that the damage threshold for STEs deposited on these substrates was above $2.92\pm0.02\,\mu$J~cm$^{-2}$.

\begin{figure}[h!]
\centering\includegraphics[width=\linewidth]{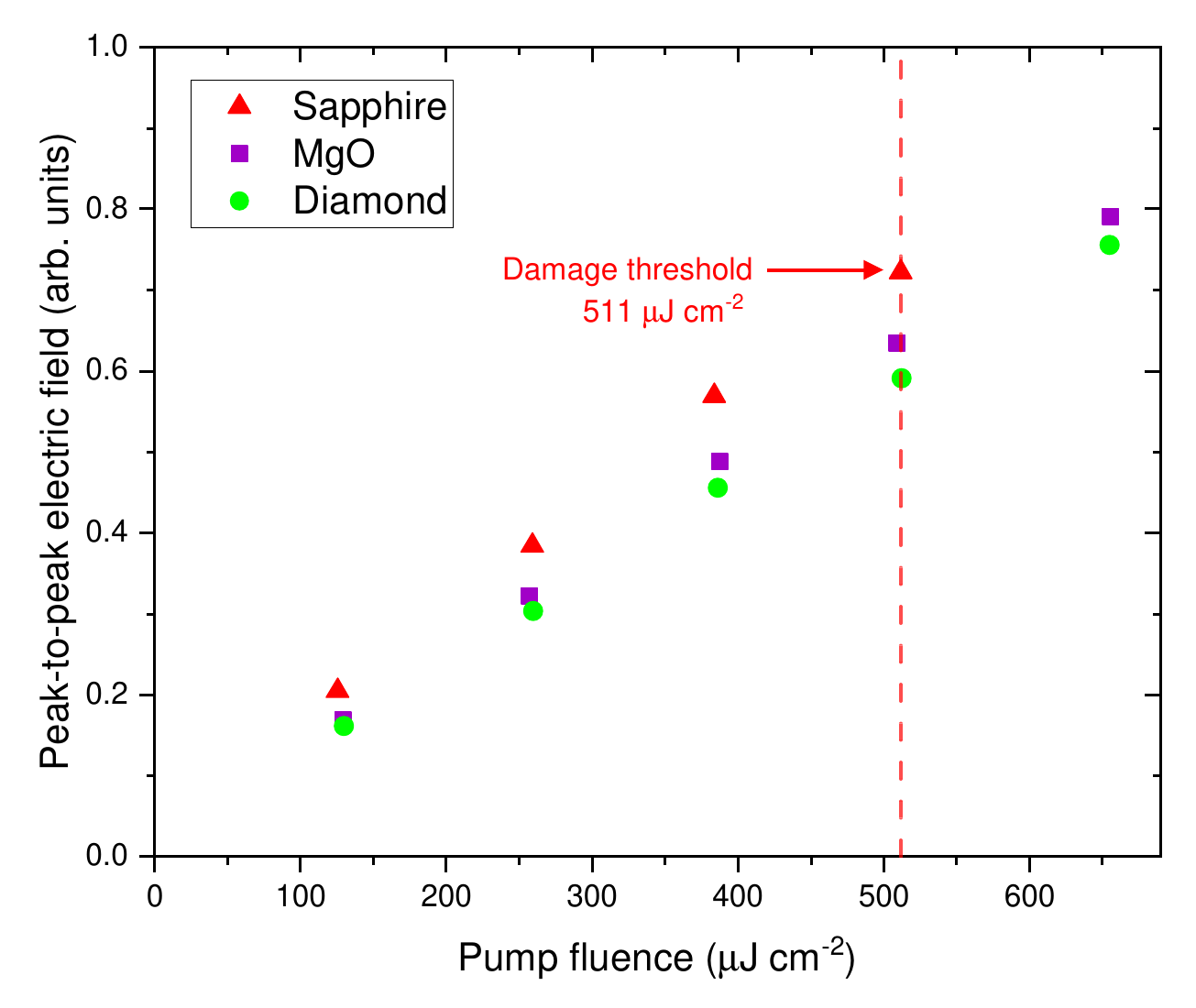}
\caption{Measured peak-to-peak THz electric field as a function of drive beam fluence for STEs grown on three substrate materials: sapphire, MgO and diamond. Measurements were taken over a high fluence range which was achieved using a spot size of $(6.0\pm0.2\,\times\,7.7\pm0.3)\,\mu$m.}
\label{fig:small_p2p}
\end{figure}

Figure \ref{fig:small_p2p} shows the the peak-to-peak THz electric field extracted from THz waveforms as a function of excitation fluence for STEs deposited on sapphire, MgO, and diamond using a smaller driving beam with dimensions $(6.0\pm0.2\,\times\,7.7\pm0.3)\,\mu$m. The highest excitation fluence applied to the STE on sapphire that did not result in damage was $0.51\pm0.02$\,mJ\,cm$^{-2}$. In this region, further fluence steps were taken, identifying a damage threshold between $0.55\pm0.02$\,mJ\,cm$^{-2}$ and $0.58\pm0.02$\,mJ\,cm$^{-2}$, and is included in Fig. S3. (Supplemental 1, Section 3). No damage is evident for the STEs on MgO or diamond, indicating that the damage thresholds are above $0.67\pm0.03$\,mJ\,cm$^{-2}$, corresponding to an increase of $1150\pm80$ times that obtained for the STE on fused silica. This indicates that using MgO or diamond substrates allows for the operation of STEs up to the recommended $\sim0.7$\,mJ\,cm$^{-2}$ where the emission dependence on fluence transitions from linear to sub-linear. \cite{Vogel2022}

\begin{figure}[h!]
\centering\includegraphics[width=\linewidth]{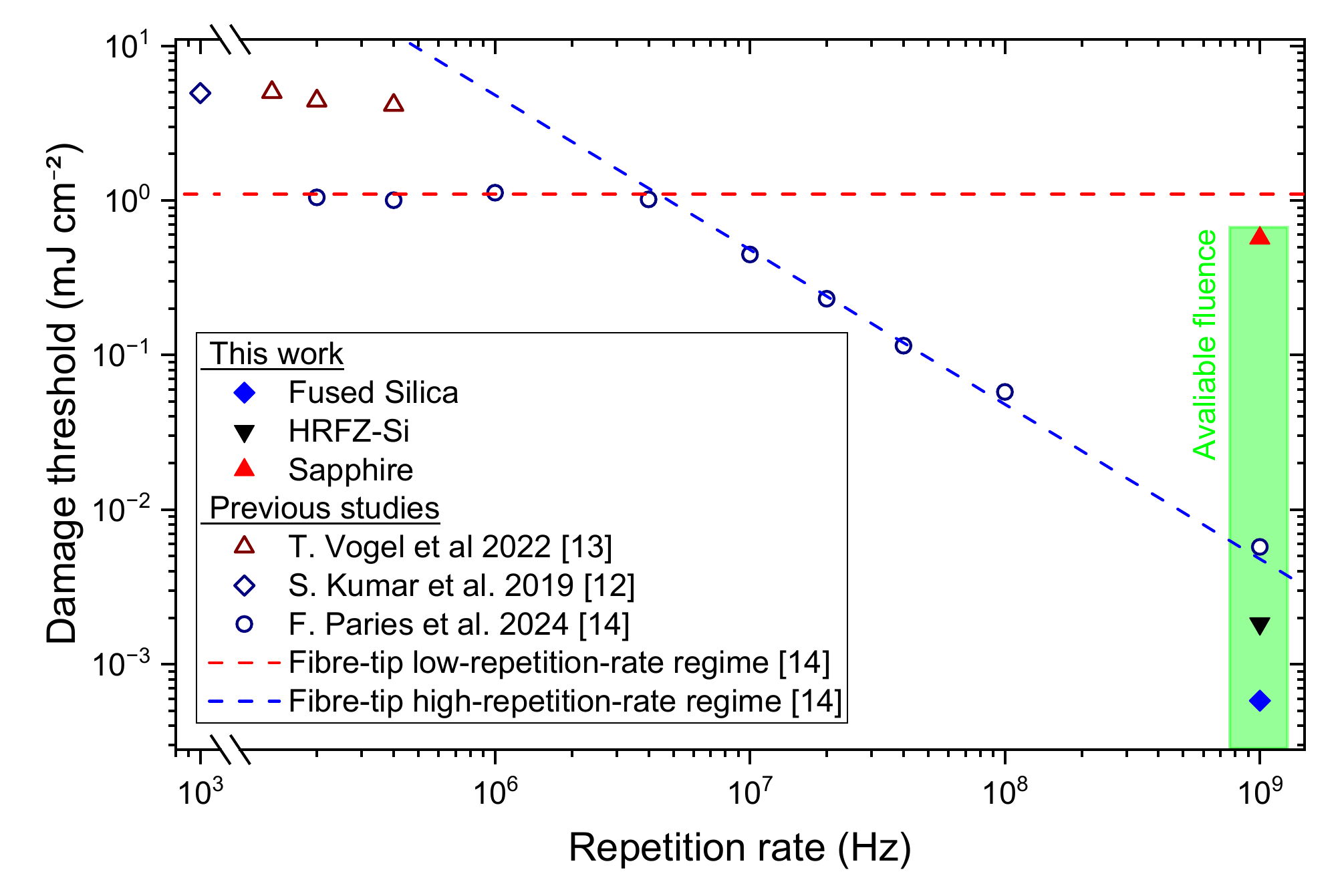}
\caption{Damage threshold verses driving laser repetition-rate from this work and reported in the literature for STEs. The shaded region indicates the range of laser drive fluences demonstrated in this work with the damage thresholds for the STEs grown on diamond and MgO falling above this region.}
\label{fig:lit_rev_comp}
\end{figure}

Figure \ref{fig:lit_rev_comp} shows the damage thresholds obtained here with values from published literature over a range of drive laser repetition rates. The high and low repetition rate regimes reported by Paries \textit{et al.} \cite{Paries2024} for STEs on fiber tips are shown as a guide for the reported damage behavior of STEs. Comparing the damage thresholds for STEs on fused silica and sapphire, the threshold is almost three orders of magnitude higher for the STE on sapphire. As damage thresholds were not obtained for STEs on MgO and diamond, these are omitted from the figure, as they are shown to have damage thresholds that exceed the maximum fluence achievable in this study. These results indicate that the thermal conductivity of the substrate is important when working at higher repetition rates to prevent thermal damage from occurring at low to moderate excitation fluences. The thermal conductivity of MgO (50\,W\,m$^{-1}$\,K$^{-1}$)\cite{MgO_K,MgO_Sp_k} and diamond (2000\,W\,m$^{-1}$\,K$^{-1}$)\cite{diamond_k} are higher than that of sapphire  (33 - 43\,W\,m$^{-1}$\,K$^{-1}$). \cite{Sp_K,Qz_Sp_K,MgO_Sp_k}. A notable exception to this is HRFZ-Si which has a reported thermal conductivity between MgO and diamond (142\,W\,m$^{-1}$\,K$^{-1}$)\cite{silicon_k} but only improves the damage threshold of the STE on fused silica by a factor of three. It is proposed that this is due to the high-reflectivity dielectric coating used to reflect the residual driving laser beam and increase the energy absorbance. \cite{Rouzegar2023,Yang2026} The coating consists of SiO$_x$ and TiO$_x$, which likely impedes the heat transfer from the STE to the HRFZ-Si. This is comparable to reports found for antenna coupled STEs grown on Si with a $0.3\,\mu$m layer of SiO$_x$ between the metal layers and the HRFZ-Si substrate. \cite{Nandi2019} However, the high reflectivity dielectric coating is necessary as it reduces the number of free carriers generated that would otherwise attenuate the emitted THz radiation. \cite{Rouzegar2023} 

\textbf{Conclusions and outlook.} In conclusion, we have demonstrated for the first time that emission from a trilayer STE is achievable using a GHz repetition rate laser. We have investigated the impact of substrate material when driven at high repetition rates and demonstrate the importance of selecting a substrate with a high thermal conductivity when attempting to combine high repetition rates with mJ\,cm$^{-2}$ fluences, which is important for use in future generations of THz-TDS systems. By utilizing substrates with high thermal conductivities, the damage threshold has been raised by at least three orders of magnitude and approaches the well-defined limit of STE performance at low-repetition rates.

\begin{backmatter}
\bmsection{Funding} United Kingdom Engineering and Physical Sciences Research Council [Grant Nos. EP/S033688/1 and UKRI1237]. 



\bmsection{Disclosures} The authors declare no conflicts of interest.

\smallskip

\bmsection{Data availability} The data that support the findings of this study are openly available in Zenodo at https://doi.org/xx.xxxxx/xxxxx.xxxxxxxxx, Ref.\cite{GethingDataset}.

\bmsection{Supplemental document}
See Supplemental 1 for the knife-edge profiles of the excitation laser beams, further details about the deposition of STEs, x-ray reflectivity measurements used to determine STE layer thicknesses and the further results about the damage for the STE grown on sapphire.
\end{backmatter}


\bibliography{References}


\end{document}